# New Results from an old Friend: The Crab Nebula and its Pulsar

**Martin C. Weisskopf** [1]

*NASA/ Marshall Space Flight Center*
*ZP 12*
*320 Sparkman Drive*
*Huntsville, AL 35801*
*USA*
*E-mail:* `martin.c.weisskopf@nasa.gov`

We summarize here the results, most of which are preliminary, of a number of recent observations of the Crab nebula system with the Chandra X-Ray Observatory. We discuss four different topics: (1) The motion on long (> 1yr) time scales of the southern jet. (2) The discovery that pulsar is not at the center of the projected ring on the sky and that the ring may well lie on the axis of symmetry but appears to be displaced at a latitude of about 5° (Note that this deprojection is by no means unique.) (3) The results and puzzling implications of the Chandra phase-resolved spectroscopy of the pulsar when compared to observations of pulse-phase variations at similar and dissimilar measures in other regions of the spectrum. (4) The search for the X-ray location of the site of the recently-discovered γ-ray flaring. We also comment briefly on our plan to use the Chandra data we obtained for the previous project to study the nature of the low-energy flux variations recently detected at hard X-Ray energies.

---

[1] Speaker

1. **Introduction**

Over the years the Crab Nebula and its pulsar have provided observers a remarkable astrophysics laboratory which continues to produce new puzzles and encourage insights as observational accuracy has improved and wavelength coverage has expanded. For a complete historical perspective up to the time of its publication, the reader is referred to the review by Hester[1]. Subsequently, there have been a number of new developments. Here we discuss four recent results from the Chandra X-Ray Observatory. Since we have a number of different collaborators involved in each of these projects, each section begins with an acknowledgement to those colleagues.

2. **Motion of the Southern Jet**

My collaborators on this project are A.F. Tennant, S.L. O'Dell, D. A. Swartz, G.G. Pavlov, and K. Mori .

Figure 1 shows two Chandra images of the Crab system taken ~8 years apart and demonstrates that the jet to the south-east has evolved with time. For example, the tail of the jet points in a somewhat different direction and the number of "kinks" (suggestive of a corkscrew) has changed. Presumeably, there is similar behavior in the northern jet, but this is more difficult to discern as that jet is masked by the radiation from the synchrotron nebula. We are currently completing a detailed study of all of the relevant Chandra observations of the southern jet to quantitiatively characterize the time evolution of changes in the mean position, the width, and the spectrum of the jet as as a function of distance from the pular and of time.

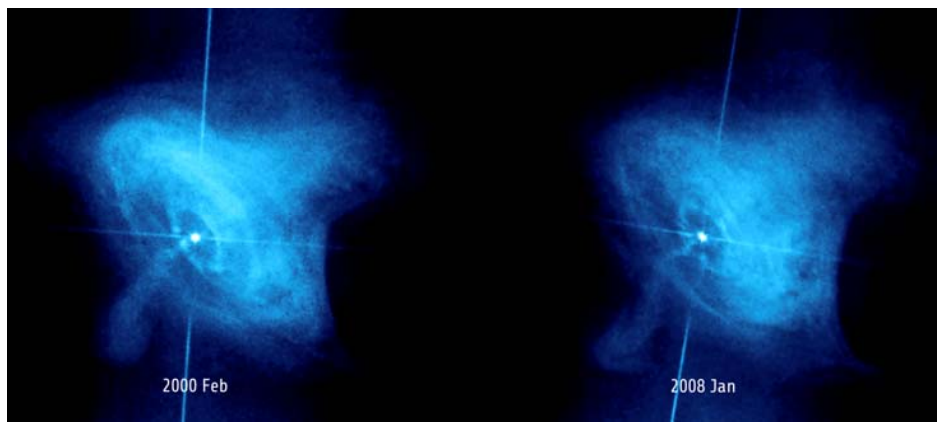

**Figure 1. Chandra Low Energy Transmission Grating zero order images of the Crab system taken in Feb 2000 (left) and Jan 2008 (right). North is up and east to the left. The line that extends mostly N-S is the grating-dispersed spectrum of the pulsar. The orthogonal line is the crossed-dispersed spectrum produced by support bars in the LETG grating facets.**

3. **The Pulsar is not necessarily at the Center of the Inner Ring**

My collaborators on this project are R.F. Elsner, J. Kolodziejczak, S.L. O'Dell, and A.F. Tennant.

The Chandra images resolve the detailed structure of its "inner ring", most likely a termination shock where pulsar-accelerated relativistic particles begin to emit X-rays [2,3]. We have analyzed such images and we find that the center of the apparent ellipse (Fig 2) —presumably a circular ring in projection—lies ~ 0.9″ (10 light-days at 2 kpc) from the pulsar's image, at a position angle of ~300° (East of North). This analysis also measures properties of the ellipse: The position angle of the semi-major axis is ~210° (East of North); and the aspect ratio 0.49.

A consequence of this discovery, i.e. that the pulsar is not symmetrical with respect to the "inner ring", is that that in a simple—albeit, not unique—de-projection of the observed geometry, a circular ring *is* centered on the axis of symmetry of the pulsar wind nebula. In this deprojection, however, although the ring is on the axis of symmetry, it is not on the equatorial plane of the pulsar, but rather lies at +4.5° latitude in pulsar-centered coordinates (Fig 2.).

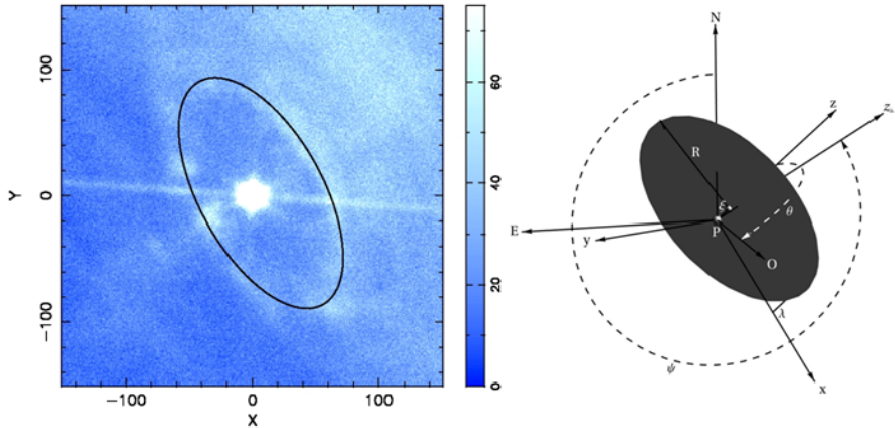

**Figure 2. Left: Chandra LETG/HRC-S zeroth-order image of the Crab Nebula (ObsIDs 758 and 759) over a 40″×40″ field centered on the pulsar. The oval line marks the peak brightness of the ellipse, which is obviously *not* centered on the pulsar. North is along the +Y axis; East, along the –X axis. The units are HRC pixels (0.132"). Right: Possible de-projected geometry. A ring of radius R is oriented to, and centered on, the symmetry axis z of the pulsar wind nebula. However, the ring lies in a plane displaced an axial distance $\xi$ from the pulsar at "P", corresponding to a latitude $\lambda = \tan^{-1}(\xi/R)$. The projection $z_\perp$ of the symmetry axis onto the sky lies at an angle $\psi$ East of North. The direction to the observer "O" is at an angle $\theta$ from the pulsar's axis of symmetry z.**

## 4. Phase resolved spectroscopy of the pulsar at all phases

My collaborators on this project are A. F. Tennant, D.G. Yakovlev, A. Harding, V.E. Zavlin, S.L. O'Dell, R.F. Elsner, and W. Becker.

In 2004[4] we presented the first Chandra-LETGS phase-resolved X-ray spectroscopy of the Crab Pulsar. More recently we made a far more efficient observation, the details of which are about to be published.[5] We show in Fig 3 one of the most interesting consequences of this research. Fig 3 compares the variation of the X-ray spectral index measured with Chandra to a number of other different parameters measured in very different spectral ranges as a function of pulse phase. The fact that the soft X-ray and hard gamma-ray spectra [6] are part of two seemingly different radiating components, and thus most likely have different emission mechanisms, raises the question of why their spectral index variation with phase should be so similar. They could share a common property, such as the same radiating particles or the same locations in the magnetosphere. This in itself poses a challenge to the theorists.

We also note the change of behavior in the X-ray power law index just before the rise of the light curve to primary pulse maximum (phases 0.83 to 0.95), the hint of a similar behavior in the γ-ray spectral index variation [6], and the abrupt change of optical polarization and position angle [7] in this same phase interval. These would appear to be correlated phenomena and thus present a further challenge to theorists.

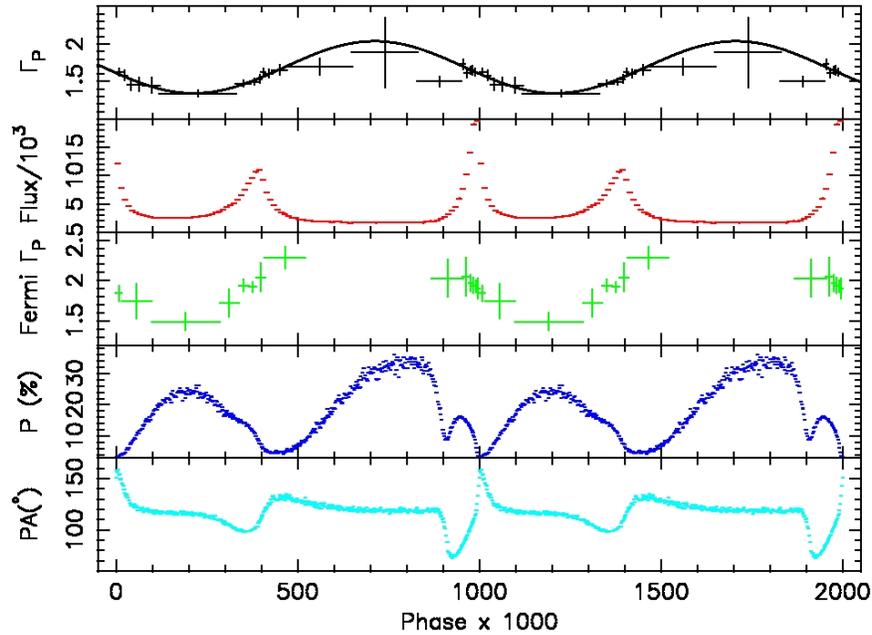

**Figure 3. Upper: the measured variation of the X-Ray powerlaw index. The next panel is the X-ray light curve (background not subtracted). Middle: the variation of the gamma-ray spectral index [6] Bottom: the two panels show the variation of the optical degree of polarization and position angle [7]. One pulse cycle is repeated twice for clarity.**

5. **Search for the origin of the γ-ray flares**

The collaborators on this project, led by A.F. Tennant, are M. Tavani, E. Costa and other members of the AGILE team, R. Buehler, R. Blandford and other members of the Fermi-Lat Team, D. Horns, C. Ferrigno and other members of the Crab Team.

Another new finding in recent times was the discovery [8,9] using the AGILE and Fermi satellites, of γ-ray flaring from the Crab Nebula. Flaring was detected at energies from 100 MeV up to ~1 GeV at high significance. A subsequent episode in April of 2011 [10,11,12] was even more intense and reached record levels. All the γ-ray observations showed that the flaring emission was not pulsed. Since radio pulsar observations did not find any evidence for a glitch[13], the flaring flaring was not directly associated with the pulsar.

Chandra was used to look at the observations prior to, and during, the flaring in April (Fig 4.) in an attempt to localize the region of the nebula wherein the flare arose.

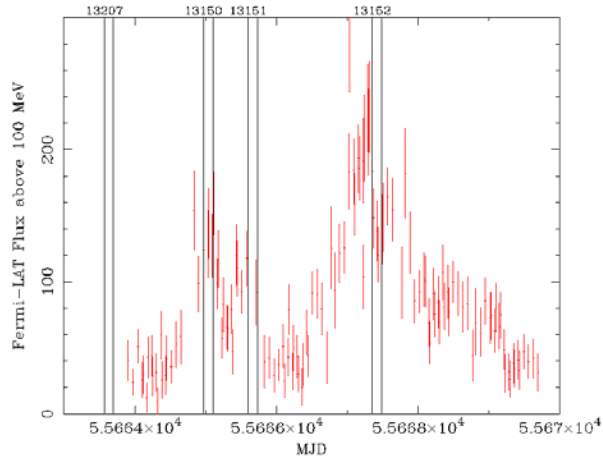

**Figure 4. Fermi-LAT light curve of the flaring emission of the Crab Nebula in 2011, March and April [14]. The vertical lines bound the times of four simultaneous Chandra observations with their ObsIDs listed at the top of the figure.**

Figure 5 shows a representative Chandra ACIS image. Also shown are a number of light curves derived from the Chandra observations that begin in September, 2010, subsequent to the announcement of the original discoveries. The closely spaced cluster of points in these light curves are from 2011, April. We emphasize that all of the X-ray light curves, not only those shown, exhibit some level of variability, and none that may be obviously associated with the γ-ray flaring. Thus, there is no clear "smoking gun". Of course, depending on the details of the emission mechanism(s), there is no guarantee that any X-ray signature must precisely coincide with the γ-ray flaring. However, that any model that requires a coincidence could/should be suspect. We hasten to add that, because of pileup in the CCDs, the pulsar itself and some of the region including and very close-by (on the scale of an arcsec or so) cannot be observed. This is why the pulsar appears as a hole in Fig. 5.

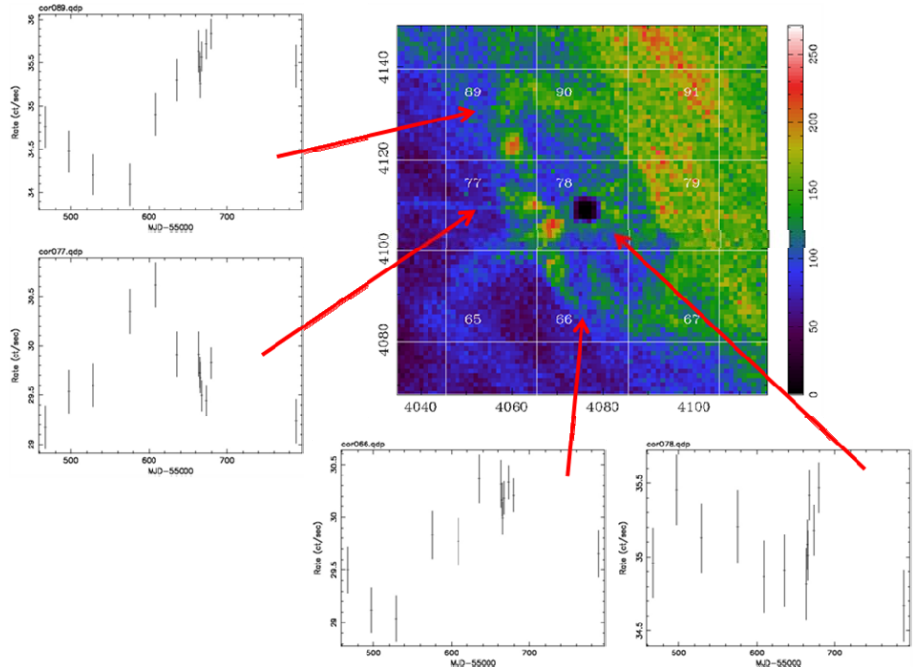

**Figure 5. Chandra image of the Crab Nebula with the pulsar appearing as a hole due to pileup in the ACIS CCD camera. The four light curves shown are discussed in the text. Image courtest A. Tennant.**

Finally, we mention the variability of the integrated X-ray emission reported by Wilson-Hodge et al. [15]. We are using the monitoring observations of the Crab that we began in late 2010, and which will continue for at least two years, to study and characterize the soft X-ray variations.

**References**

[1] Hester, J. J. 2008, ARA&A, **46**, 127
[2] Rees, M. J., & Gunn, J. E. 1974, MNRAS, **167**, 1
[3] Kennel, C. F., & Coroniti, F. V. 1984, ApJ, **283,** 694
[4] Weisskopf, M.C. et al. 2004, ApJ, **605**, 360
[5] Weisskopf, M.C. et al. arXiv:1106.3270
[6] Abdo, A. A. et al. 2010, ApJ, **708**, 1254
[7] Slowikowska, A., Kanbach, G., Kramer, M. & Stefanescu, A. 2008, in HIGH TIME RESOLUTION ASTROPHYSICS: The Universe at Sub-Second Timescales. AIP Conference Proceedings, **984**, 51
[8] Tavani, M. et al. 2011, Science, **331**, 736
[9] Abdo, A.A. et al. 2011, Science, **331**, 739
[10] Buehler, R. D'Ammando, F. & Cannon, A. 2011, ATel 3276.
[11] Hays, E., Buehler, R. & D'Ammando, F. 2011, ATel 3284
[12] Tavani, M., Bulgarelli, A. , Striani, E., et al. 2011, ATel 3282
[13] Espinoza, C.M. et al. 2010, ATel 2889.
[14] Courtsey R. Buehler
[15] Wilson-Hodge, C. et al. 2011, ApJ, **727,** L40